\documentclass[secnumarabic,amssymb, nobibnotes, aps, prl, preprint]{revtex4-1}
\usepackage{graphicx}
\usepackage{amsmath}
\usepackage[caption=false]{subfig}

\pdfoutput=1

\begin{document}
\title{Stable Switching among High-Order Modes \\in Polariton Condensates}
\author{Yongbao Sun,${}^1{}^\ast$  Yoseob Yoon,${}^1$ Saeed Khan,${}^2$  Li Ge,${}^{3,4}$ Loren N. Pfeiffer,${}^2$ Ken West,${}^2$ Hakan E.  T$\ddot{\textrm{u}}$reci,${}^2$ David W. Snoke,${}^5$ and Keith A. Nelson${}^1{}^\ast$}
\affiliation{${}^1$Department of Chemistry and Center for Excitonics, Massachusetts Institute of Technology, 77 Massachusetts Avenue, Cambridge, MA 02139, USA \\
${}^2$Department of Electrical Engineering, Princeton University, Princeton, NJ 08544, USA\\
${}^3$Department of Engineering Science and Physics, College of Staten Island, City University of New York, New York 10314, USA\\
${}^4$The Graduate Center, College of Staten Island, City University of New York, New York 10016, USA\\
${}^5$Department of Physics, University of Pittsburgh, 3941 O'Hara St., Pittsburgh, PA 15218, USA\\
}
\begin{abstract}{We report multistate optical switching among high-order bouncing-ball modes (``ripples'')  and whispering-gallerying modes  (``petals'') of exciton-polariton condensates in a laser-generated annular trap.  By tailoring the diameter and power of the annular trap, the polariton condensate can be switched among different trapped modes, accompanied by redistribution of spatial densities and superlinear increase in the emission intensities, implying that polariton condensates in this geometry could be exploited for a multistate switch. A model based on non-Hermitian modes of the generalized Gross-Pitaevskii equation reveals that this mode switching arises from competition between pump-induced gain and in-plane polariton loss. The parameters for reproducible switching among trapped modes have been measured experimentally, giving us a phase diagram for mode switching. Taken together, the experimental result and theoretical modeling advances our fundamental understanding of the spontaneous emergence of coherence and move us toward its practical exploitation.}
\end{abstract}
\maketitle

\noindent\textbf{Introduction.}  Strong coupling between cavity photons and excitonic resonances of a quantum well (QW) placed inside the cavity leads to the formation of new mixed eigenstates known as exciton-polaritons (hereafter simply polaritons). They behave as bosons with extremely low effective mass and overall repulsive interactions when excitation densities are low. The photonic and excitonic fractions can be varied by adjusting the relative detuning of photon and exciton resonances, typically by varying the cavity width in a wedged sample structure. This allows direct control over the polariton-polariton interaction strength, which increases with the excitonic fraction. For a review of polariton properties, see SI and Ref.~\cite{Kavokin2007}.

Polaritons provide a unique testbed for the study and manipulation of macroscopic quantum effects. Quantum phenomena such as Bose-Einstein condensation have been reported from liquid helium temperature \cite{Kasprzak2006, Balili2007, thermalization} up to room temperature \cite{Christopoulos2007, Plumhof2014, Kena-Cohen2010} in various systems. This not only allows the investigation of quantum phenomena at elevated temperatures in a convenient fashion, but also presents exciting opportunities to create all-optical polaritonic devices. As a consequence, great efforts have been devoted to the development of techniques for  manipulating the properties of microcavity polaritons \cite{Balili2007, Idrissi-Kaitouni2006,Lai2007, Kim2011,Cerda-Mendez2010,Cristofolini2013, Askitopoulos2013, Dreismann2014, Askitopoulos2015} .
 
Previous experiments on Bose condensation of polaritons were usually performed with the photonic resonance close to the excitonic resonance, which resulted in highly excitonic characteristics in polaritons. Together with short cavity lifetimes, this severely limited the distance polaritons could propagate \cite{Manni2011, Askitopoulos2013, Askitopoulos2015}. The development of new structures with much longer cavity photon lifetimes, from 20--30 ps \cite{Wertz2010} to well over 100 ps \cite{Nelsen2013, Steger2015, Liu2015}, has allowed the possibility of polariton propagation over macroscopic distances. This property was recently used to measure the polariton-polariton interaction strength in a region with no free excitons \cite{interactions}. 

In the present work, we generated polaritons with high photonic fractions by choosing a region of the wedged sample with large negative cavity detuning. Their highly photonic nature allowed the polaritons to propagate coherently over long distances to form condensate states with radial extent up to 100 $\mu$m inside an optical trap formed by an annular pattern of excitation light. While interactions of  polaritons in this case are not strong enough for them to thermalize into an equilibrium gas, they still play an important role. The interactions of polaritons with excitons in the pump region allow the polaritons to undergo Bose condensation inside the optical trap. Furthermore, nonlinear polariton-polariton interactions result in switching among different condensate modes at high pump powers. The spatial distributions of these modes vary dramatically with very small changes of the excitation densities, but are temporally very stable as long as the excitation power is stable. This stability has allowed us to map out the phase boundaries between different modes in our optical trap.  Upon state switching as the excitation power is increased, the emission intensities from the condensates also increase in a superlinear fashion. The large changes not only allow us to experimentally distinguish different quantum states, but also strongly suggests the use of polaritons in multistate switching applications.


\vspace*{0.3cm}
\noindent\textbf{Petal and ripples in the annular trap.} Annular-shaped beams with diameters ranging from 42 $\mu$m to 107 $\mu$m were used to excite a high-$Q$ microcavity structure that has a cavity lifetime of $\sim$135 ps. This allows polaritons to propagate over macroscopic distances of up to millimeters \cite{Steger2013, Steger2015}. The laser beam was tuned about 140 meV above the bandgap of the QW material; therefore it essentially generated free carriers, which subsequently relaxed down to exciton and polariton states (see Methods for experimental details). Petals and ripples were formed inside the excitation annulus, with radial extent up to 100 $\mu$m. In theory, if not limited by the pump power, higher-order condensate states with length scales on the order of millimeters could be realized in this high-$Q$ microcavity structure, making them entirely visible by eye. 


\begin{figure}[htbp]
\centering
\includegraphics[width=1.\textwidth]{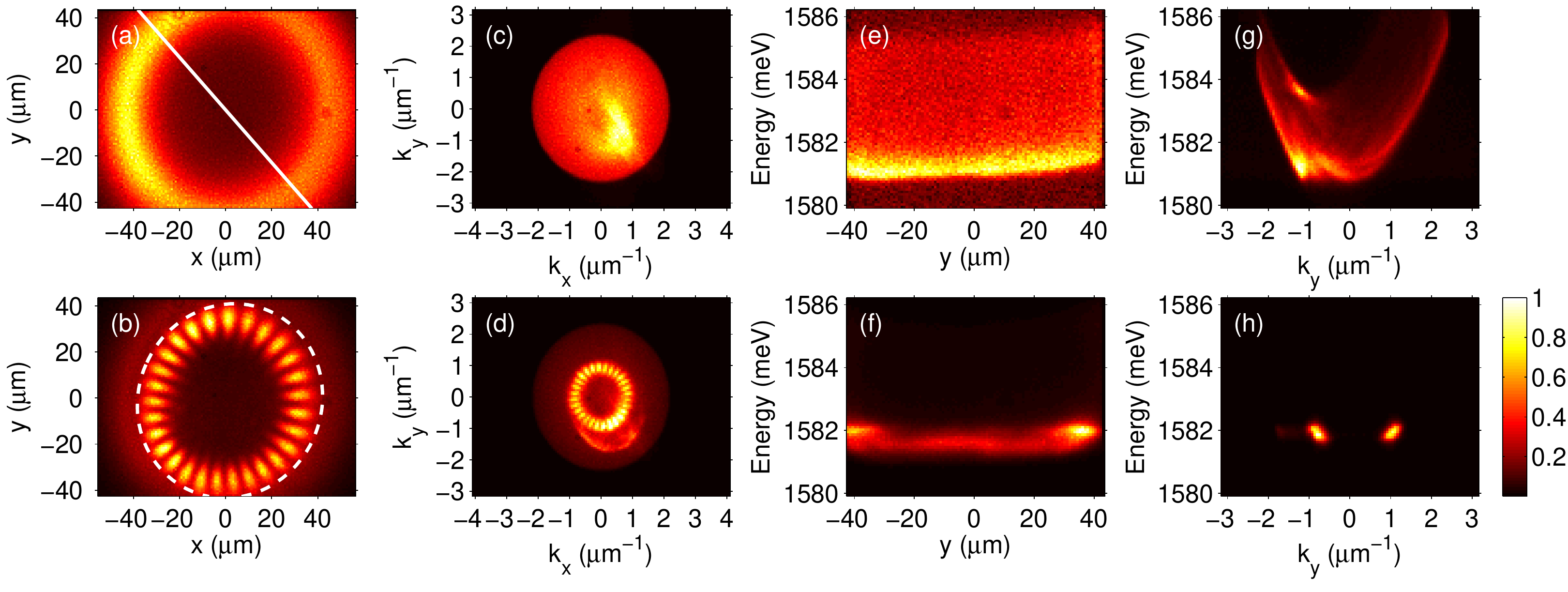}
\caption{ (color online). {\bf Petals in the annular trap with a diameter of 82 $\mu$m}. (a)-(b) polariton density distribution below (a) and above (b) condensation threshold under annular-shaped beam excitation. (c)-(d) Polariton momentum distribution below (c) and above (d) condensation threshold. (e)-(f) Energy-resolved polariton density distribution at $x = 0$ below (e) and above (f) condensation threshold. (g)-(h) Energy-resolved polariton momentum distribution at $k_x=0$  below (g) and above (h) condensation threshold. The white solid line in (a) indicates the direction of the cavity energy gradient (photon energy decreasing from upper right to lower left), and the white dashed line in (b) shows the position of the annular pump.}
\label{petals}
\end{figure}

Petals are whispering-gallery modes in the annular trap. Fig.~\ref{petals} shows the emission patterns from an annular trap with a diameter of 82 $\mu$m. Polaritons remain in the vicinity of the pump region below the condensation threshold, as shown in Fig.~\ref{petals}a. The asymmetry in the density distribution is largely due to inhomogeneity within the pump intensity profile. Fig.~\ref{petals}c shows the momentum distribution of the polaritons below the condensation threshold. Because the photonic mode in the microcavity we used has an energy gradient of $\sim$11 meV/mm along the white  line in Fig.~\ref{petals}a, there is a net flow of polariton fluid along this energy gradient, as evidenced by the accumulation of the polariton densities with in-plane wavevector components at $k_x\approx1\,\mu$m$^{-1}$ and $k_y\approx-1\,\mu$m$^{-1}$ in Fig.~\ref{petals}c. The cavity gradient can also be identified from the energy-resolved emission profile in Fig.~\ref{petals}e at low pump powers. In this plot, the $x=0$ $\mu$m slice of Fig.~\ref{petals}a was projected onto the entrance slit of the spectrometer CCD and then spectrally dispersed.  As can be seen, there is an energy difference of $\sim$0.5 meV between the emissions at $x=\pm40$ $\mu$m. The propagation effect can also be identified in the energy-resolved $k$-space emission profile as a smeared dispersion, which has been reported in Ref.~\cite{Nelsen2013}  with the same sample structure.

When the excitation density is above the condensation threshold, polaritons propagate over 10 $\mu$m toward the center of the trap and form the petal state inside the excitation ring. The position of the pump annulus is plotted in Fig.~\ref{petals}b as white dashed lines, along with the emission profile from petals. The petals demonstrate nodal structures similar to those of the high-order whispering-gallerying modes in lasers, with the periodicity matching the density accumulation at $k_x=1$ $\mu$m$^{-1}$ and $k_y=-1$ $\mu$m$^{-1}$. The petal structure is also observed in momentum space as expected since the condensate is a coherent state and the density distributions in position space and momentum space are Fourier-transform related. As expected, the energy-resolved measurements show far narrower emission spectra from the condensates (f and h) than from polaritons below the condensation threshold (e and g). Above the threshold, petals typically have higher energies than those polaritons that have flowed to the center of the annular trap, as seen in Fig.~\ref{petals}f. 


\begin{figure}[hdtp]
\centering
\includegraphics[width=1.\textwidth]{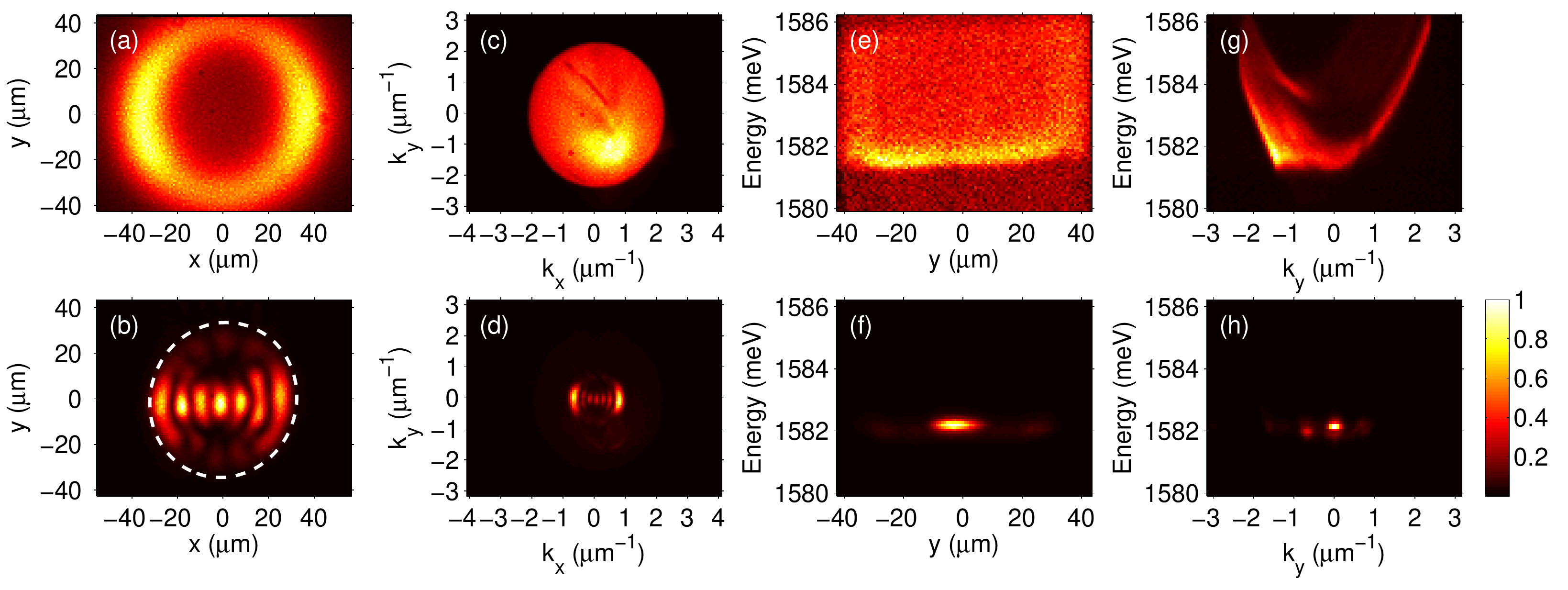}
\caption{ (color online). {\bf Ripples  in the annular trap with a diameter of 66 $\mu$m.} (a)-(b) polariton density distribution below (a) and above (b) condensation threshold under annular-shaped beam excitation. (c)-(d) Polariton momentum distribution below (c) and above (d) condensation threshold. (e)-(f) Energy-resolved polariton density distribution at $x = 0$ below (e) and above (f) condensation threshold. (g)-(h) Energy-resolved polariton momentum distribution at $k_x=0$  below (g) and above (h) condensation threshold. }
\label{ripples}
\end{figure}

Unlike petals, ripples are radially confined bouncing-ball modes in the annular trap. In Fig.~\ref{ripples}a, we plot the emission profiles observed when a 66-$\mu$m annulus was used to excite the microcavity. Below the condensation threshold, the distributions of polaritons in real and momentum space show very similar signatures to those in the previous case. However,  confined ripples appear when the excitation density is above the condensation threshold, as shown in Fig.~\ref{ripples}b. Similar patterns have been studied in quantum chaotic systems where they were termed as bouncing-ball modes \cite{McDonald1988}. In $k$-space, we observed two large populations of polaritons at $k_x=\pm$1 $\mu$m$^{-1}$ indicative of a ripple mode, together with several states with smaller but not negligible amount of polaritons. This suggests that the ripple pattern in Fig.~\ref{ripples}b arises from the interference of these paired momentum states. Figures \ref{ripples}f and h show energy-resolved emission along the vertical slices $x=0$ and $k_x = 0$ in Fig.~\ref{ripples}b and d accordingly. Again the emission spectra narrow dramatically above the condensation threshold.

In this work, the higher-order condensate states appear at a lower threshold than the lowest-order condensate state at $k_{||}=0$, unlike the case in Ref.~\cite{interactions}, where polaritons are composed of higher fractions of excitons. This confirms that interactions play a very important role in the formation of the lowest-order versus higher-order condensate states. In particular, a balance of polariton leakage from the pump region against amplification from the reservoir determine whether ripples or petals will define the lowest-threshold mode; this is expanded upon in the theory section.



\begin{figure}[htbp]
\centering
\includegraphics[width=.75\textwidth]{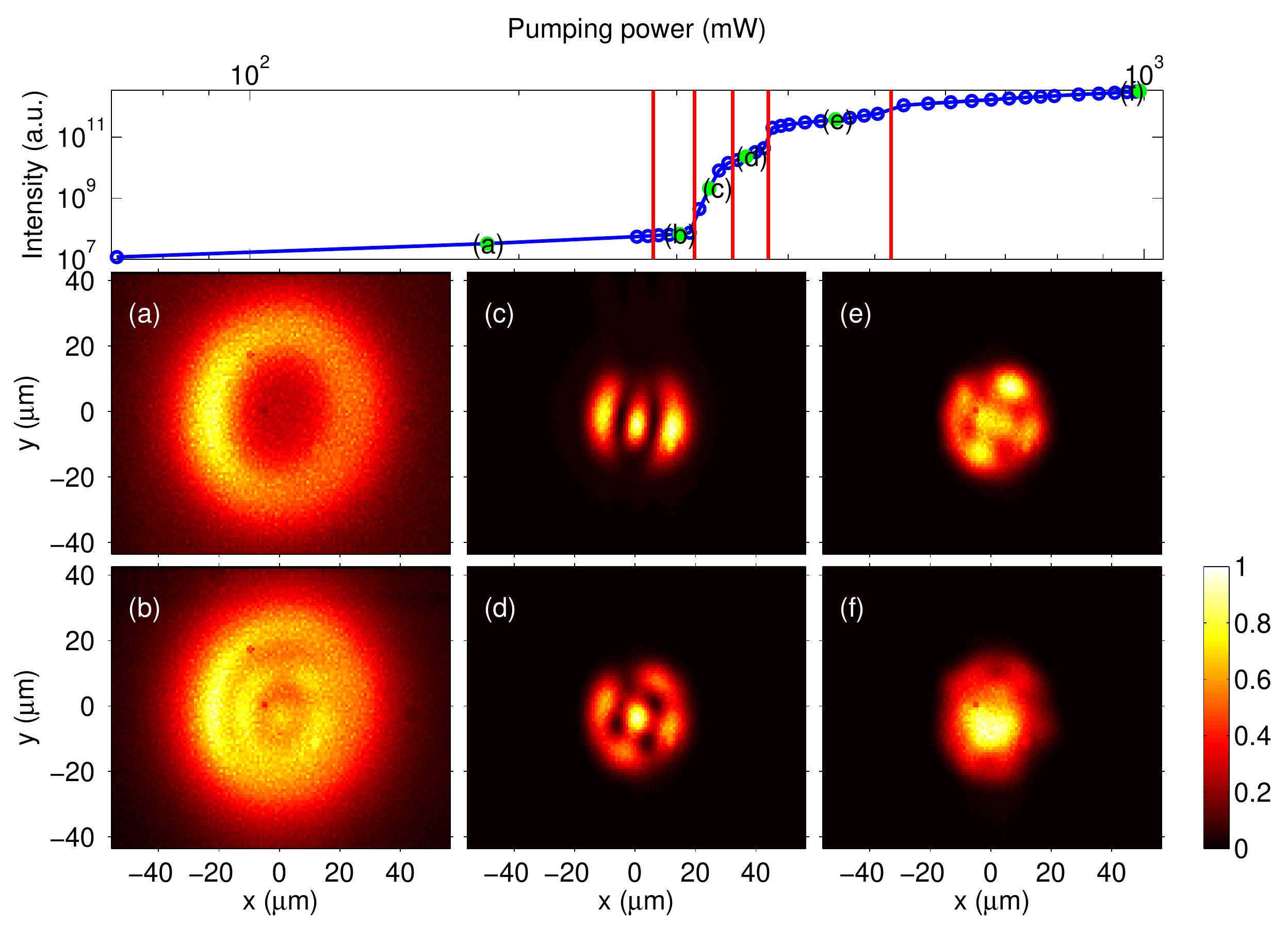}
\caption{ (color online). {\bf Mode switching in a 42 $\mu$m trap.} Top: the PL intensity as a function of pump power. The red line indicates boundaries of different quantum states. The green dots are selected pump power levels for which the normalized real-space density distributions of the quantum states are  shown in (a)-(f). }
\label{switch}
\end{figure}

\vspace*{0.3cm}
\noindent\textbf{Stable mode switching in condensates.} The condensate can be switched among various petal and ripple states by varying the pump power continuously. In the top panel of Fig.~3, we show the integrated emission intensity in the field of view as a function of pump power. The intensity undergoes several distinct sharp jumps, which are marked by the red lines, and increases by five orders of magnitude when the pump power is increased by a factor of only $\sim$15. The real-space density distributions corresponding to the green dots in the upper panel are shown in Fig.~\ref{switch}a-f. We clearly identify that the jumps in emission intensity are accompanied by redistributions of the real-space densities, that is, by mode switching.

In Fig.~3a, the excitation level was still below the condensation threshold, and patterns similar to those in Fig.~1a and ~2a were observed. Figure~3b demonstrates the onset of a higher-order state, but it was very difficult to resolve reliably. In Fig.~3c, a two-node ripple mode appears. Figure 3d and e are  mixtures of both petals and ripples. Numerical simulations discussed below suggest that petals and ripples coexist at this power due to  interactions between these states. As shown in Fig.~3f, when the system was pumped very hard, all the higher-order quantum states collapsed to the lowest-order condensate state. This power tunability of mode switching not only allows us to distinguish different high-order modes, but also suggests that polariton condensates in the annular trap can be implemented in device applications for a stable multistate switch. With better control of the pump power, we believe more states can be accessed independently.

\begin{figure}[htbp]
\centering
\subfloat{\includegraphics[width=0.42\textwidth]{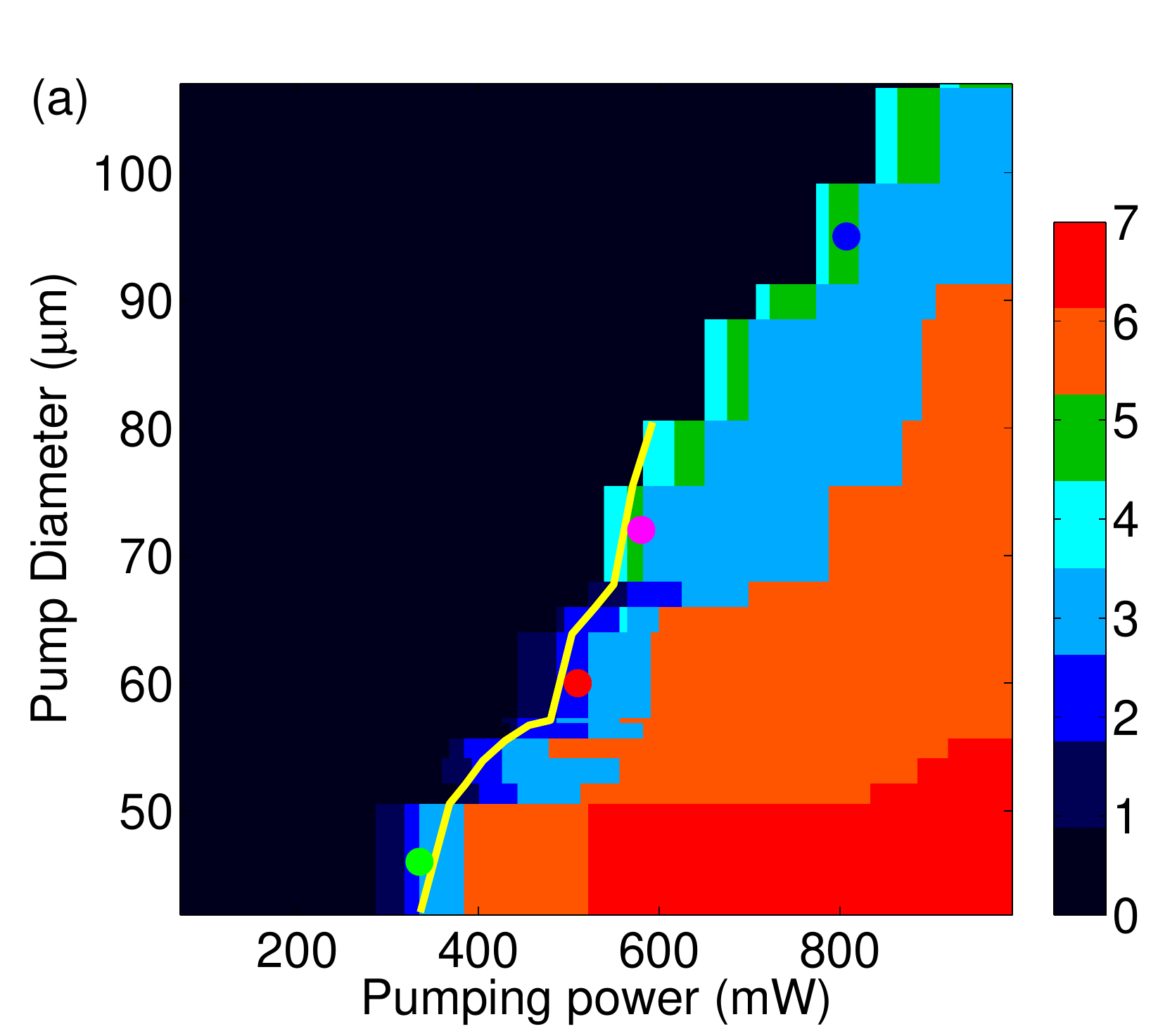}} 
\subfloat{\includegraphics[width=0.42\textwidth]{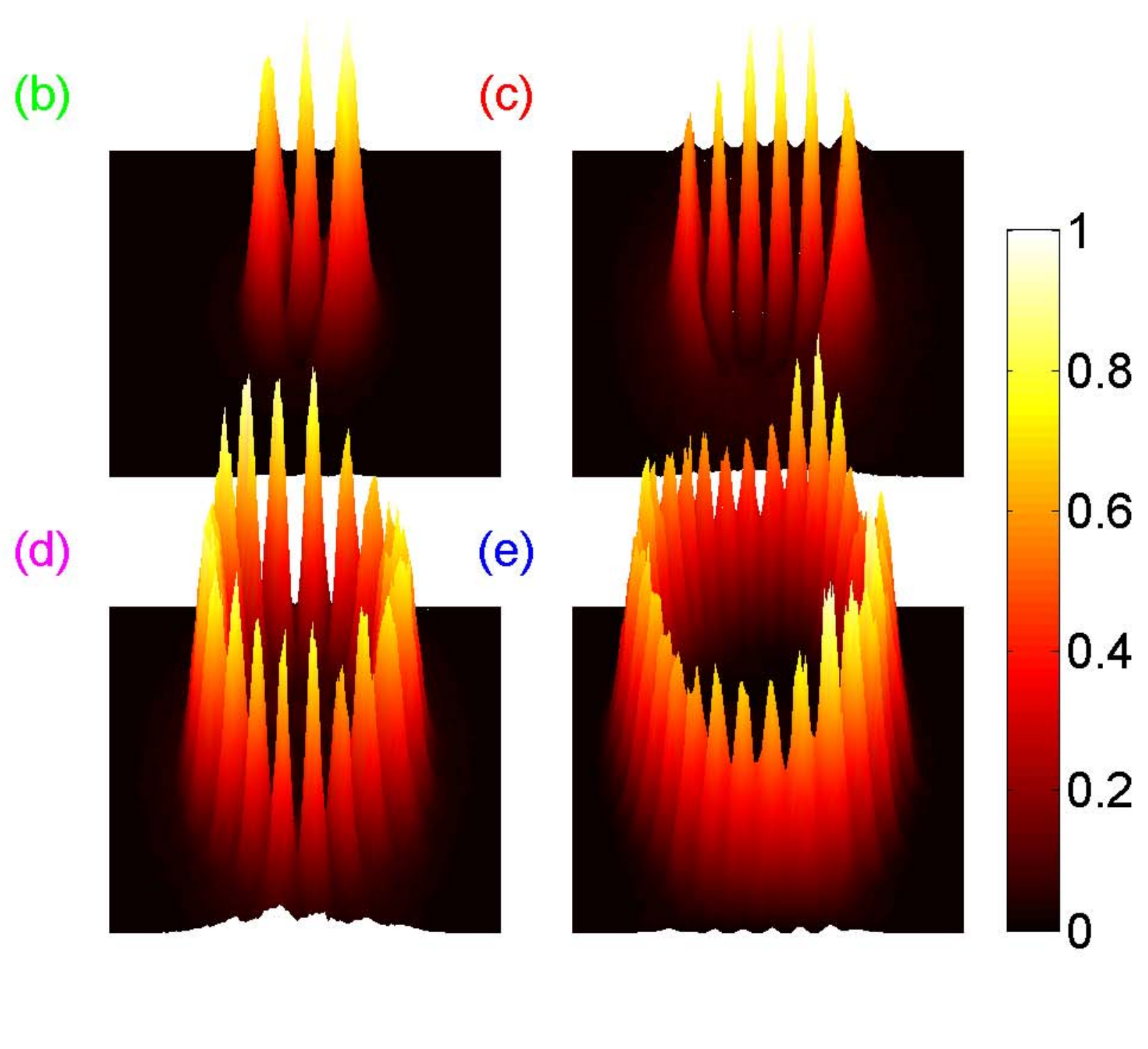}} 
\caption{ (color online). {\bf Phase boundaries of different quantum states in the annular trap.} (a) Measured phase boundaries with annular pumps. Yellow solid line is the simulated boundary for the lowest-threshold mode. Different colors correspond to different types of states, with blue, green and red indicating ripples,  petals and lowest-order condensate states. The black-shaded region is uncondensed polaritons, and the other colors are mixtures of modes similar to those in Fig.~3d and e. Spatial distributions of modes indicated by colored dots are shown in (b)-(e) correspondingly. }
\label{phase_diagram}
\end{figure}

\vspace*{0.3cm}
\noindent\textbf{Phase boundaries of higher-order quantum states.} In order to fully characterize the phase boundaries between different quantum states, we recorded the real-space polariton density distributions with excitation ring diameters ranging from 42 $\mu$m to $107$ $\mu$m and pump power ranging from 50 mW to 1 W. Because of the stability of the distributions and the superlinear increase in the emission intensities as shown in Fig.~\ref{switch}, we were able to classify different quantum states at different pump conditions. The resulting phase boundaries are shown in Fig.~\ref{phase_diagram}. In this plot, different colors are assigned to different types of states with distinct spatial distributions. The black-shaded region (0) in the upper left region indicates the uncondensed polaritons. Blue (2) and green (5) stand for ripples and petals, respectively. Both petals and ripples exist in a very narrow range of the phase diagram. This indicates that switching among polariton condensate states in the optical trap is very sensitive and reliable. The different regions of the staircase structure of the phase diagram show modes with different numbers of nodes that were observed with incremented values of the excitation ring size and power. The lowest-order condensate states, coded as red (7),  occupy the lower right region of the phase map. The rest of the colors indicate patterns that are mixtures of high-order states, similar to those shown in Fig.~\ref{switch}d and e (See SI for more spatial distributions of these mixed modes). 

Based on this phase diagram, we can see that as the excitation density and ring size are increased, ripples and petals appear successively as the lowest-threshold modes, and the phase boundary for the lowest-threshold modes is approximately linear. Both features will be explained in the following theory section.  The number of lobes in either petals or ripples can be easily tuned by changing the pump parameters, as shown in Fig.~4b-e. This measured phase boundary should serve well to calibrate the implementation of an exciton-polaritonic multistate switch by making use of the high-order quantum states.

\begin{figure}[t!]
\begin{center}
\includegraphics[scale=0.6]{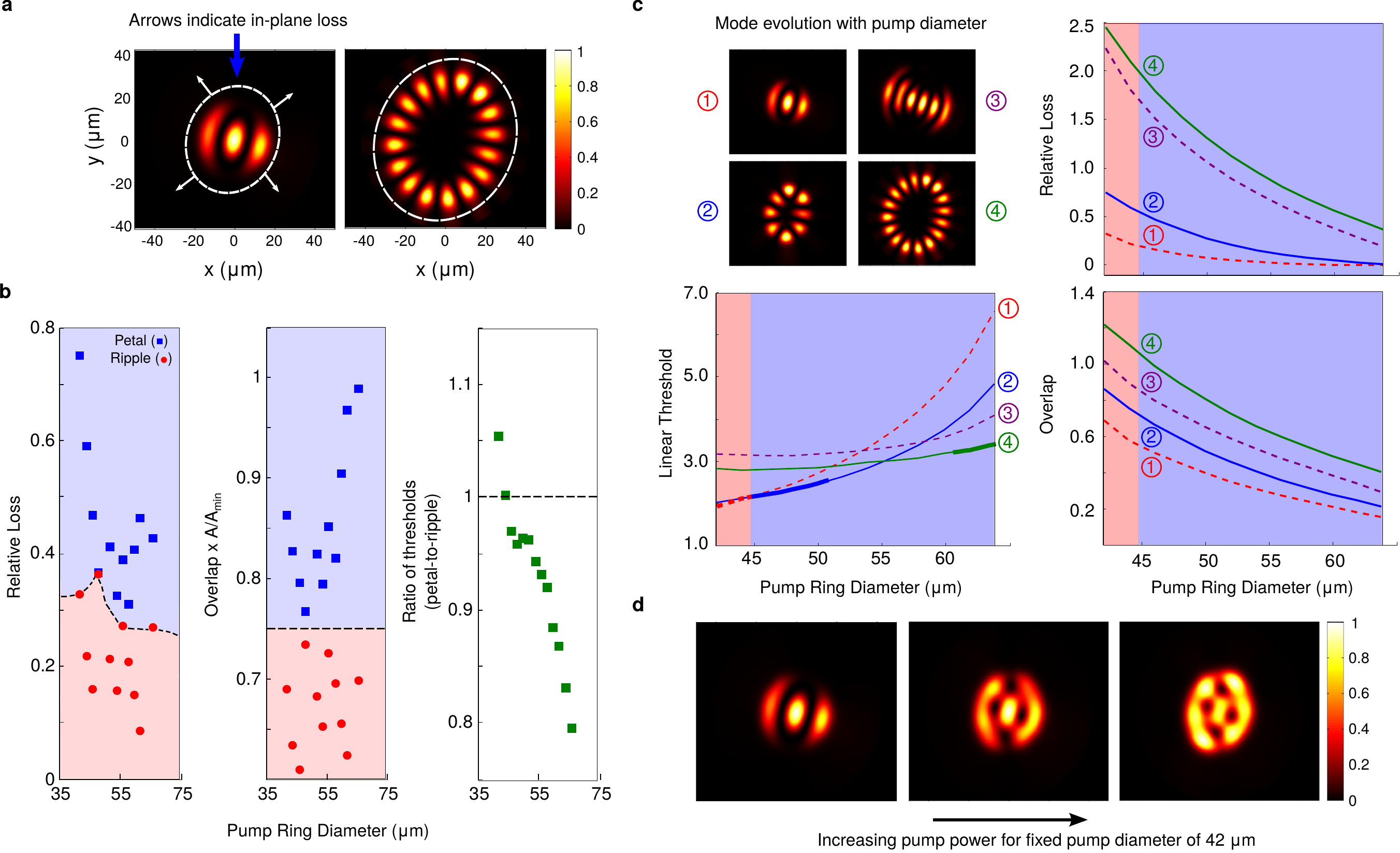}
\caption{\textbf{Numerical simulations.} (a) Lowest threshold modes for pump diameters $42~\mu$m and $67~\mu$m respectively; white dashed lines indicate the outer pump edge. (b)  Loss-overlap characteristics for \emph{lowest threshold} petal and ripple modes, and petal-to-ripple threshold ratio, against pump diameter. Overlap is scaled by a dimensionless factor $\propto$ pump area $A$ to highlight the separation between ripple and petal modes. (c) Loss, overlap, and threshold evolution for two ripple modes (dashed lines 1 and 3) and petal modes (solid lines 2 and 4) shown in the top panel. Red (blue) shaded area denotes pump diameters where ripple (petal) modes have lowest threshold. A thick line in the threshold plot indicates the lowest threshold mode; in the region with no thick lines, a mode other than those considered here has lowest threshold. (d)  Simulation of condensate density for increasing pump power at a fixed pump diameter of $42~\mu$m.}
\label{lossOverlap}
\end{center}
\end{figure}

\vspace*{0.3cm}
\noindent\textbf{Theory and numerical simulation of pattern formation.} Below threshold, reservoir-condensate dynamics for incoherently pumped polaritons can be described using an effective Gross-Pitaevskii equation (GPE) linearized in the condensate density (for details, see SI):
\begin{align}
i\frac{\partial \Psi}{\partial t} &= \mathcal{H}_{\rm L}(P)\Psi = \left[-\frac{\nabla^2}{2m} + \frac{g_R}{\gamma_R}Pf(\mathbf{r}) + \frac{i}{2} \left(\frac{R}{\gamma_R}Pf(\mathbf{r}) - \gamma_c\right)  \right] \Psi.
\label{linGPE}
\end{align}
where $\mathcal{H}_{\rm L}(P)$ is the linear, \emph{non-Hermitian} generator of condensate dynamics: it describes polariton decay (rate $\gamma_c$) and gain (rate $R$) through stimulated scattering from the exciton reservoir generated by the pump (profile $f(\mathbf{r})$, strength $P$), in addition to the real-valued reservoir-condensate repulsion ($\propto g_R$). When the pump power is weak, the condensate density reflects quantum fluctuations and is almost zero.  Beyond a power $P_n^{\rm th}$, the $n$th eigenmode $\varphi_n$ of $\mathcal{H}_{\rm L}(P_n^{\rm th})$ becomes an unstable fluctuation around the uncondensed steady state, corresponding to a condensate mode with frequency given by the real part of its eigenvalue. By varying the pump power, a set of such spatial modes $\{\varphi_n(\mathbf{r};P_n^{\rm th},\omega_n)\}$ can be obtained, with linearized power thresholds $\{P_n^{\rm th}\}$ and real frequencies $\{\omega_n\}$. This linearization is exact until condensation first occurs, and thus the linearized mode with lowest threshold is especially significant: it is the actual mode first observed upon condensation. Naturally, the following question arises: what determines the spatial mode with lowest condensation threshold? Using a continuity equation for the condensate density derived from the GPE, we arrive at a simple formula for the \emph{linearized} threshold $P_n$ for condensation of the $n$th mode~\cite{Ge2013}:
\begin{align}
\frac{P_{n}^{\rm th}}{P_0} = \frac{ 1 + \gamma_n/(\rho_n\gamma_c)}{ \mathcal{G}_n} \equiv \frac{1+ \Gamma_n}{\mathcal{G}_n}~;~P_0 = \frac{\gamma_c\gamma_R}{R}
\label{thresholdFormula}
\end{align}
where
\begin{align}
\rho_n = \int_{\mathcal{P}}d^2\mathbf{r}~|\varphi_n(\mathbf{r})|^2~,~\mathcal{G}_n = \frac{1}{\rho_n}\int_{\mathcal{P}}d^2\mathbf{r}~f(\mathbf{r})|\varphi_n(\mathbf{r})|^2~;~\gamma_n = \oint_{\partial\mathcal{P}} \vec{j} \cdot d\vec{s}~,~\vec{j} = \frac{i}{2m}\left( \varphi_n\vec{\nabla}\varphi_n^* - c.c.\right)
\label{lossOverlapExps}
\end{align}
For a given mode, the threshold is determined by: (i) relative loss $\Gamma_n$, which compares in-plane loss $\gamma_n$ to total mirror loss $\rho_n\gamma_c$, the former being the flux of probability current $\vec{j}$ leaking across the \emph{outer} pump edge $\partial\mathcal{P}$ (see Fig.~\ref{lossOverlap}a), and (ii) $\mathcal{G}_n$, a dimensionless measure of the overlap between the mode and the pump within the region $\mathcal{P}$ enclosed by this pump edge. The lowest threshold mode minimizes Eq.~(\ref{thresholdFormula}) by maximizing overlap with the pump to benefit from amplification, while still having low density near $\partial \mathcal{P}$ to reduce the relative loss $\Gamma_n$. Note that as relative loss for a mode becomes smaller, its overlap becomes increasingly more important in determining the threshold.

We study the linear modes of $\mathcal{H}_{\rm L}(P)$ for a range of pump diameters; the loss-overlap characteristics for the \emph{lowest threshold} ripple and petal modes are plotted in Fig.~\ref{lossOverlap}b. While petal modes have stronger overlap and higher loss than ripple modes, their threshold decreases below that of ripple modes as pump diameter increases. To understand this, we focus for clarity on one low order and one higher order mode for petals (labeled 2 and 4) and ripples (1 and 3), and consider their loss, overlap, and threshold evolution as a function of pump diameter in Fig.~\ref{lossOverlap}c. pump, we see that relative loss decreases with increase in pump diameter; physically, it reflects the increase of the quality factor of the condensate modes, which in turn is due to a better confinement and shorter tail of the condensate modes in the radial direction, leading to reduced in-plane loss.  Due to this effect, there exists for each mode a large enough pump diameter at which its relative loss is small enough such that the overlap $\mathcal{G}_n$ primarily determines its threshold; in this competition, petal modes have an advantage over ripple modes. Therefore, below a critical pump diameter petals are typically too lossy to have lower thresholds than ripple mode, even though their pump overlap is stronger. Beyond this critical diameter, $\Gamma_n$ for petals decrease enough for their stronger overlap to pull its threshold down below that of the competing ripple mode. For annular profiles, a transition diameter will always exist due to this decrease of $\Gamma_n$; the particular diameter depends on details of the profile. By extension, for higher order states with higher relative loss (see Fig.~\ref{lossOverlap}c), larger pump diameters are needed than those for lower order states until $\Gamma_n$ decreases sufficiently to encourage condensation into these modes, in agreement with observations here. Finally, we note that overlap $\mathcal{G}_n$ decreases with growing pump diameter since the pump density $Pf(\mathbf{r})$ goes down as $1/R$ for a radius $R$ annular pump with fixed FWHM. From Eq.~(\ref{thresholdFormula}) and (\ref{lossOverlapExps}), the resultant decrease in $\mathcal{G}_n$ increases $P_n^{\rm th}$ linearly with pump diameter; this is apparent from the simulated lowest threshold boundary, in good agreement with the experimental phase diagram shown in Fig.~\ref{phase_diagram} .

Going beyond the condensation threshold requires full simulation of the nonlinear GPE over a large spatio-temporal grid; the unprecedentedly large condensate sizes (upto $\sim$$100~\mu$m) observed in the current work, together with polariton wavelengths ($\sim$$1~\mu$m) demanding fine spatial (and hence temporal) resolution make such simulations very computationally expensive here. We circumvent this issue by expanding the condensate wavefunction in a \emph{pump power-dependent}, non-Hermitian basis set $\{\varphi_n(\mathbf{r}; P,\omega_n)\}$ that account for the spatial complexity of the linearized condensate problem, with time-dependent coefficients. For discrete values of $P=P_n$, one mode in each set reduces to the corresponding threshold modes introduced before (note that in general $\omega_n$ is complex, but for that particular threshold mode it is on the real axis).   This reduces the full nonlinear GPE and reservoir dynamical equation to a set of coupled ODEs, an effective nonlinear coupled mode theory for reservoir-condensate dynamics (details in a future publication~\cite{Khan2015}). Applying to the specific case of a pump of diameter $42~\mu$m, the coupled mode theory reveals mixing of lowest threshold modes beyond threshold, when polariton-polariton interactions within the condensate become important; in particular, the coexistence of petal and ripple states shown in Fig.~\ref{lossOverlap}d was reproduced using this theory.

\vspace*{0.3cm}
\noindent\textbf{Conclusion and outlook.} 
We have seen the stable formation of high-order quantum states, including ripples, petals and their coherent mixtures, under non-resonant excitation, with a well-defined phase diagram in the pump parameter space. Ripples are confined bouncing-ball modes while petals are whispering-gallery modes in the trap. The all-optical trapping allows facile switching among these condensate states in the annular trap, accompanied by superlinear increases in the emission intensities.

The measured patterns bear some similarity to the multiple modes seen in standard vertical-cavity, surface-emitting lasers (VCSELs), e.g. the petal patterns seen in Ref.~\cite{Li2012}. However, in typical lasers and VCSELs, the system hops uncontrollably between different modes, leading to unwanted noise (e.g., Ref.~\cite{Pedaci2005}). The nonlinear interactions in the polariton condensate system stabilize the modes to resist multimode behavior. This means that this system acts effectively as multistable optical switch, in which transitions between states can be effected by small changes of the input light beam. 

\vspace*{0.3cm}
\noindent\textbf{Experimental methodologies.} {\footnotesize  The microcavity used in this work is GaAs based structure grown by molecular beam epitaxy. The cavity has an exceptionally high qualify factor of $\sim$320,000, which corresponds to a polariton lifetime of $\sim$270 ps at resonance. During the experiment, the sample was thermally attached to a cold finger in an open-loop cryostat which was stabilized at 10 K. The excitation laser is a commercial continuous-wave (c.w.) laser, and was modulated by an acousto-optic modulator at 1 kHz with a duty cycle of 0.5\% to prevent unwanted sample heating. The annular trap was generated by shaping the phase front of the c.w. laser using a high-resolution spatial light modulator. Because of the eccentricity in the pump profile, which is approximately 0.3, the diameters reported here are geometric means of the lengths of major and minor axes of the pattern. The photoluminescence of polaritons was collected in a reflection geometry using an objective lens with a numerical aperture of 0.28, and was relay imaged to a spectrometer CCD. The energy-resolved emissions were obtained by spectrally dispersing a specific slice of either the far-field or near-field image selected by the entrance slit of the spectrometer CCD.}

\bibliographystyle{naturemag}
\bibliography{patterns}
\vspace*{0.3cm}
\noindent\textbf{Acknowledgements.} Y.S., Y.Y. and K.A.N were supported as part of the Center for Excitonics, an Energy Frontier Research Center funded by the US Department of Energy, Office of Science, Office of Basic Energy Sciences under Award Number DE-SC0001088. S.K. and H.E.T. were supported by the National Science Foundation under Grant Number DMR-1151810, L.G. were supported by CIRG 21 Grant from City University of New York, D.W.S. was supported by the National Science Foundation under Grant Number DMR-1104383. L.N.P. and K.W. were partially funded funded by the Gordon and Betty Moore Foundation through the EPiQS initiative Grant GBMF4420, and by the National Science Foundation MRSEC Grant DMR-1420541.

\vspace*{0.3cm}
\noindent\textbf{Author contributions} \\ Y.S. and K.A.N. designed the experiments; Y.S. and Y.Y. performed the experiment; S.K., L.G. and H.E.T. carried out the numerical simulation. Y.S. and S.K. analyzed the data; L.P.N and K.W. fabricated the microcavity structure; all the authors participated to the results discussion and manuscript preparation.

\vspace*{0.3cm}
\noindent\textbf{Additional information} \\Supplementary Information is available in the online version of the paper. Reprints and permissions information is available online at www.nature.com/reprints. Correspondence and requests for materials should be addressed to Y.S. at \href{mailto:ybsun@mit.edu}{ybsun@mit.edu} or to  K.A.N. at \href{mailto:kanelson@mit.edu}{kanelson@mit.edu}.
\vspace*{0.3cm}

\noindent\textbf{Competing financial interest} \\The authors declare no competing financial interests.
\newpage
\appendix
\renewcommand{\thefigure}{S\arabic{figure}}
\renewcommand{\theequation}{S\arabic{equation}}
\setcounter{figure}{0}    
\setcounter{equation}{0}    

\begin{center}
{\bf \large  Supplementary Information: \\ Stable Switching of Higher-Order Modes in Polariton Condensates}
\end{center}
\begin{center}
{Yongbao Sun,${}^1{}^\ast$  Yoseob Yoon,${}^1$ Saeed Khan,${}^2$  Li Ge,${}^{3,4}$\\ Loren N. Pfeiffer,${}^2$ Ken West,${}^2$ Hakan E.  T$\ddot{\textrm{u}}$reci,${}^2$ David W. Snoke,${}^5$ and Keith A. Nelson${}^1{}^\ast$}
\end{center}

\begin{center}
\fontsize{10pt}{20pt}{\it ${}^1$Department of Chemistry and Center for Excitonics, Massachusetts Institute of Technology, 77 \\ Massachusetts Avenue, Cambridge, MA 02139, USA}\par
\fontsize{10pt}{20pt}{\it ${}^2$Department of Electrical Engineering, Princeton University, Princeton, NJ 08544, USA}\par
\fontsize{10pt}{20pt}{\it ${}^3$Department of Engineering Science and Physics, College of Staten Island, City University of New York, New York 10314, USA}\par
\fontsize{10pt}{20pt}{\it ${}^4$The Graduate Center, College of Staten Island, City University of New York, New York 10016, USA}\par
\fontsize{10pt}{20pt}{\it ${}^5$Department of Physics, University of Pittsburgh, 3941 O'Hara St., Pittsburgh, PA 15218, USA}
\end{center}

\begin{figure}[htbp]
\centering
  \includegraphics[width=0.85\textwidth]{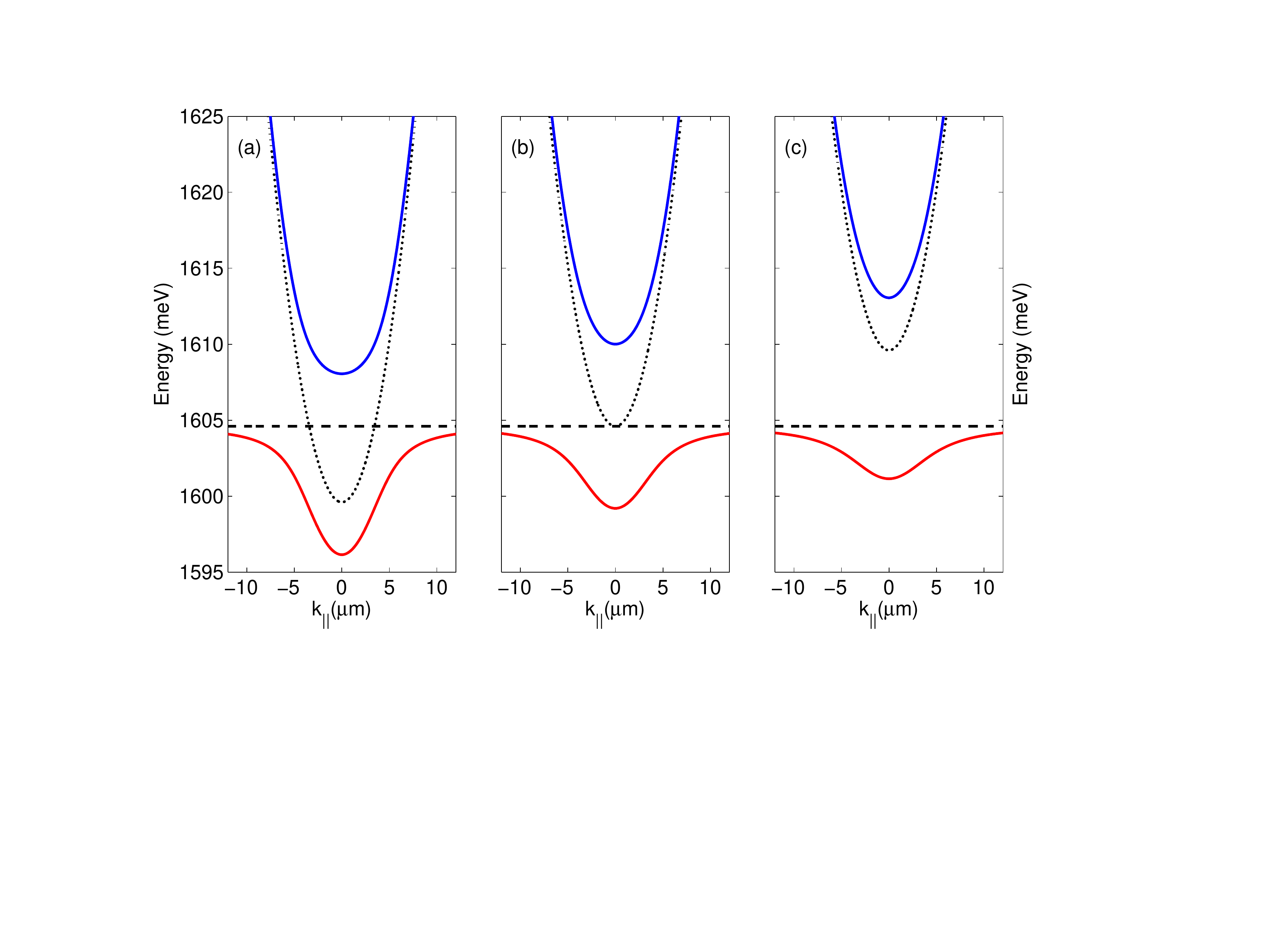}
   \caption{ (color online) Dispersion curves of polariton at three representative cavity detunings (a) $\delta = -5$ meV. (b) $\delta = 0$ meV. (c) $\delta = 5$ meV. The dotted line shows the confined cavity mode, and the dashed line shows the bare exciton mode. The blue and red solid lines indicate the upper polariton (LP) and lower polariton (UP) branches, respectively, arising from the strong coupling between corresponding cavity modes and exciton modes. Calibrated sample parameters were used in the calculations.}
\label{energy_dispersion}
\end{figure}

\vspace*{0.3cm}
\noindent\textbf{\textrm{Background on exciton-polaritons in semiconductor microcavities}}.\hspace{10pt}  Exciton-polaritons are formed in semiconductor microcavities through the strong coupling between optical modes of the microcavity and exciton transitions of material embedded inside the microcavity. For the case of a single microcavity mode and a single exciton transition, two polariton modes, the upper and lower polaritons, are formed with the energies of the two polariton modes, $E_{LP}(k_{||})$ and $E_{UP}(k_{||})$, given by:
\begin{align}
E_{LP/UP}(k_{||}) = \frac{1}{2}\left[E_{X}({k_{||}}) + E_{C}(k_{||}) \mp \sqrt{\Omega^2 + \delta^2(k_{||})}\right]
\label{eq:Epolariton}
\end{align}
where $k_{||}$ is the wave vector in the plane perpendicular to the microcavity confinement direction, $E_X(k_{||})$ is the energy of the exciton transition, $E_C(k_{||})$ is the energy of the cavity mode, $\delta(k_{||})$ is the detuning energy defined as $\delta(k_{||})=E_C(k_{||})-E_X(k_{||})$, and $\Omega$ is the strength of radiative coupling between the exciton and cavity field, also known as full Rabi splitting energy. The confinement of light gives the cavity mode a parabolic dispersion in the plane perpendicular to the confinement direction: $E_C=\hbar^2k_{||}^2/2m_C$, where $m_C$ is the effective mass of the cavity field. This effective mass is typically $10^{-4}$ times lighter than the vacuum electron mass, and about $10^{-3}$ times less than an exciton in a GaAs quantum well structure, so that $E_X(k_{||})$ is essentially constant with $k_{||}$. The energies $E_X(k_{||}),E_C(k_{||}),E_{LP}(k_{||})$, and $E_{UP}(k_{||})$ as given in Fig.~\ref{energy_dispersion} for three different values of $\delta(k_{||}=0)$. The energies were calculated using (\ref{eq:Epolariton}) and parameters matching the sample structure used in the experiments: $\Omega = 10.84$ meV and $E_X(0)=1604.6$ meV. 

\begin{figure}[htbp]
\centering
  \includegraphics[width=0.50\textwidth]{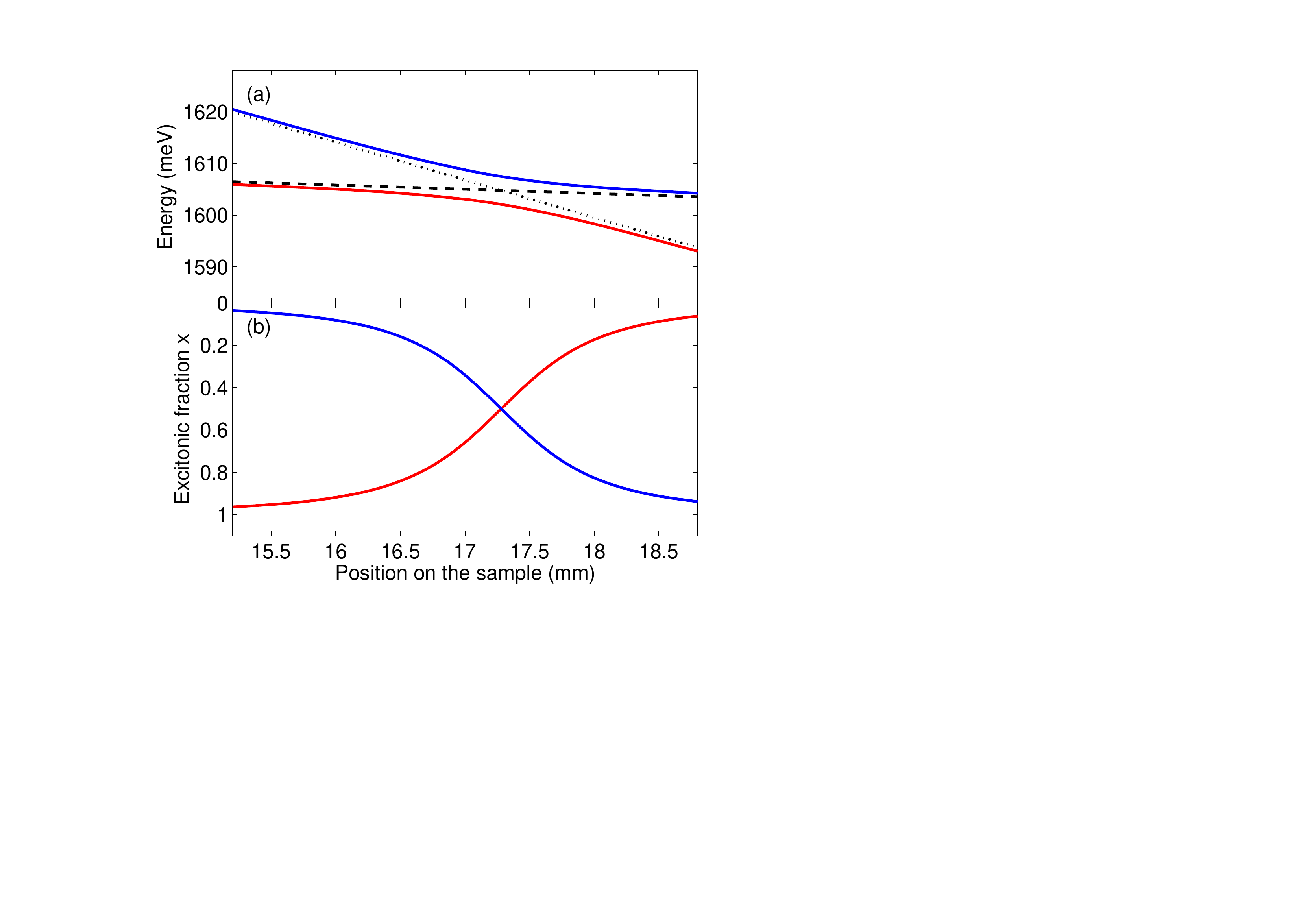}
   \caption{ (color online) (a) The calculated upper polariton (blue line) and lower polariton (red line) energies at different positions of the sample. The dashed lines indicate the exciton energies, and the dotted line shows the cavity energies. (b) Excitonic fractions of upper polaritons (blue line) and  lower polaritons (red line) at different sample positions.}
\label{spatial_dispersion}
\end{figure}

The length of the cavity increases monotonically along one direction of the QW plane so that the energy of the cavity mode can be tuned relative to the exciton resonance energy, as shown in Fig.~\ref{spatial_dispersion}, allowing us to experimentally tune $\delta(k_{||}=0)$. The energies of all modes in Fig.~\ref{energy_dispersion} are plotted as a function of $k_{||}$, the in-plane wave vector. As can be seen in this figure, $E_X(k_{||})$ is essentially constant with respect to $k_{||}$ while $E_C(k_{||})$ is parabolic.

The polariton modes are linear superpositions of the exciton and microcavity photon modes. The lower polariton and upper polariton operators, $\hat{P}_{k_{||}}$ and $\hat{Q}_{k_{||}}$, respectively, can be written in terms of exciton and cavity operators, $\hat{a}_{k_{||}}$ and $\hat{b}_{k_{||}}$:
\begin{align}
\hat{P}_{k_{||}} &= X(k_{||})\hat{a}_{k_{||}}+C(k_{||})\hat{b}_{k_{||}k_{||}}\\
\hat{Q}_{k_{||}} &=-C(k_{||})\hat{a}_{k_{||}}+X(k_{||})\hat{b}_{k_{||}}.
\end{align}
The coefficients, $X(k_{||})$ and $C(k_{||})$, are called the exciton and cavity Hopfield coefficients and are given by 
\begin{align}
|X(k_{||})|^2 &= \frac{1}{2}\left(1+\frac{\delta(k_{||})}{\sqrt{\delta^2(k_{||})+\Omega^2}}\right)\\
|C(k_{||})|^2 &= \frac{1}{2}\left(1-\frac{\delta(k_{||})}{\sqrt{\delta^2(k_{||}k_{||})+\Omega^2}}\right)
\end{align}

The characteristics of the polariton modes are determined by the coefficients, which depend on $\delta(k_{||})$. The lower polariton is more photon-like and the upper polariton is more exciton-like for $\delta(k_{||})<0$, and the lower polariton is more exciton-like and the upper polariton is more photon like $\delta(k_{||})>0$. Due to the wedge in the cavity thickness, we can easily tune the excitonic fraction $|X(k_{||})|^2$ of lower polaritons by moving the excitation spot at different positions, as shown in Fig.~\ref{spatial_dispersion}b, where we plot $|X(k_{||}=0)|^2$ at different positions on the sample. As seen in Fig.~\ref{energy_dispersion} and Fig.~\ref{spatial_dispersion}, the energies and shapes of the polariton dispersion curves depend strongly on $\delta$:
positive detuning results in lower polaritons that are more exciton-like, with a heavier effective mass and stronger interactions with phonons and other carriers, while negative detuning results in lower polaritons that are more photon-like, with a smaller lower polariton mass and weaker interactions with phonons and other carriers.

\begin{figure}[htbp]
\centering
  \includegraphics[width=0.60\textwidth]{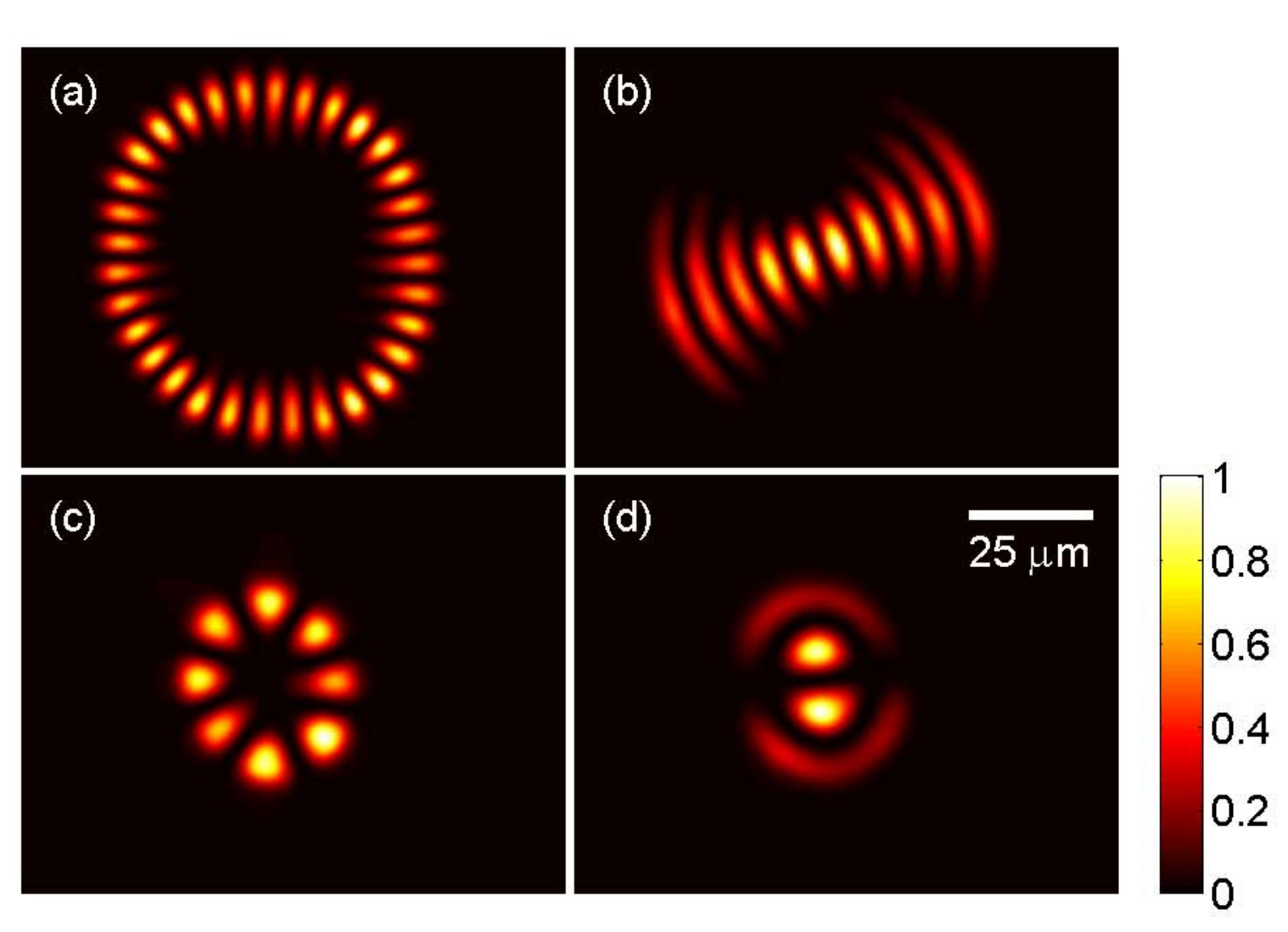}
   \caption{ (color online) (a) The calculated upper polariton (blue line) and lower polariton (red line) energies at different positions of the sample. The dashed lines indicate the exciton energies, and the dotted line shows the cavity energies. (b) Excitonic fractions of upper polaritons (blue line) and  lower polaritons (red line) at different sample positions.}
\label{simu}
\end{figure}

\vspace{.5cm}
\noindent\textbf{\textrm{Simulated petal and ripples.}}  By using the experimental pump profile as $f(\boldsymbol{r})$ in the nonlinear Gross-Pitaevskii equation, the petal and ripples observed in the experiment can be qualitatively reproduced.  Fig.~\ref{simu} shows the resulting polariton density distributions. The pump profiles that generated the patterns in
 Fig.~1(f) were used in the simulation for Fig.~\ref{simu}(a) and (b), and the number of lobes was exactly reproduced using this method. The relative intensities of the peaks along the azimuthal direction were also qualitatively captured by this model. In the simulation, a ripple state showed up at a lower energy  ($\sim$0.05 meV lower) but with significantly higher pump density ($\sim$50\%). This mode was not observed in the experiment, due to the limited pump density we can use. By using the pump profile corresponding to that of Fig.~1(e), the reproduced polariton density distribution agreed well with that observed in the experiment, shown in Fig.~\ref{simu}(c) and (d). Additionally, another two patterns were identified, and they have energies similar to the first mode (within 0.01 meV) with a difference in the pump threshold by less than 2\%. This agrees with what we saw in Fig.~4(d) and (e), where both ripple and petals were seen in the time-integrated measurements, and is also captured by our mode-integration simulation as shown in Fig.~5 in the main text.


\vspace*{0.3cm}
\noindent\textbf{Gross-Pitaevskii equation and linearization.} We use a generalized Gross-Pitaevskii equation (GPE) to describe the dynamics of microcavity exciton-polaritons under incoherent pumping. In this standard approach, the nonlinear interactions of polaritons within the condensed fraction are treated at the mean-field level, while pumping and losses are introduced as complex-valued terms, so that the generalized GPE for the dynamics of the condensate wavefunction $\Psi(\mathbf{r},t)$ has the form:
\begin{align}
i\frac{\partial \Psi}{\partial t} = \left[ -\frac{\nabla^2}{2m} + g_Rn_R + \frac{i}{2} \left(Rn_R - \gamma_c\right) \right] \Psi + g|\Psi|^2\Psi
\label{GPE}
\end{align}
where for clarity we have suppressed the $(\mathbf{r},t)$ dependence of the polariton wavefunction and the density $n_R$ of the pump-generated exciton reservoir. This reservoir gives rise to a repulsive term describing the interaction of condensate polaritons with reservoir excitons, with strength $g_R$, together with an amplification of the condensed fraction via stimulated scattering from the reservoir at rate $R$. This latter gain contribution together with the inclusion of polariton mirror loss at rate $\gamma_c$ make the effective generator describing condensate dynamics non-Hermitian in this case. Finally, the polariton-polariton repulsion within the condensate appears as the nonlinear term $\propto g$ at the mean-field level. The dynamics of the pump-induced reservoir must also be accounted for by a dynamical equation of the form:
\begin{align}
\frac{\partial n_R}{\partial t} = Pf(\mathbf{r})  - Rn_R|\Psi|^2 -\gamma_R n_R
\label{res}
\end{align}
 $P$ and $f(\mathbf{r})$ are the pump strength and spatial profile as described in the main paper, the source of the exciton reservoir. The aforementioned scattering from the exciton reservoir into the condensate at the rate $R$ causes a depletion of the reservoir, which is encapsulated in the second term on the right hand side. Reservoir losses that occur via mechanisms other than scattering into the reservoir (e.g. recombination losses) are described by $\gamma_R$.

For pumping below the condensation threshold, the system has a steady state with a pump generated exciton density and an uncondensed polariton state. The steady state reservoir density in this regime can be obtained after linearizing Eq.~(\ref{res}) by dropping nonlinear terms of order $|\Psi|^2$; in this steady-state regime the exciton reservoir density adiabatically follows the pump:
\begin{align}
n_R(\mathbf{r},t\to \infty) = \frac{P}{\gamma_R} f(\mathbf{r})
\end{align}
Below threshold, a linearization of the GPE is also valid; we can replace $n_R(\mathbf{r},t)$ by its linearized steady state value, and neglect the nonlinear polariton-polariton interactions $\propto g$. This yields the linearized GPE for condensate dynamics, Eq.~(\ref{linGPE}) of the main text. 

\begin{figure}[t]
\begin{center}
\includegraphics[scale=0.6]{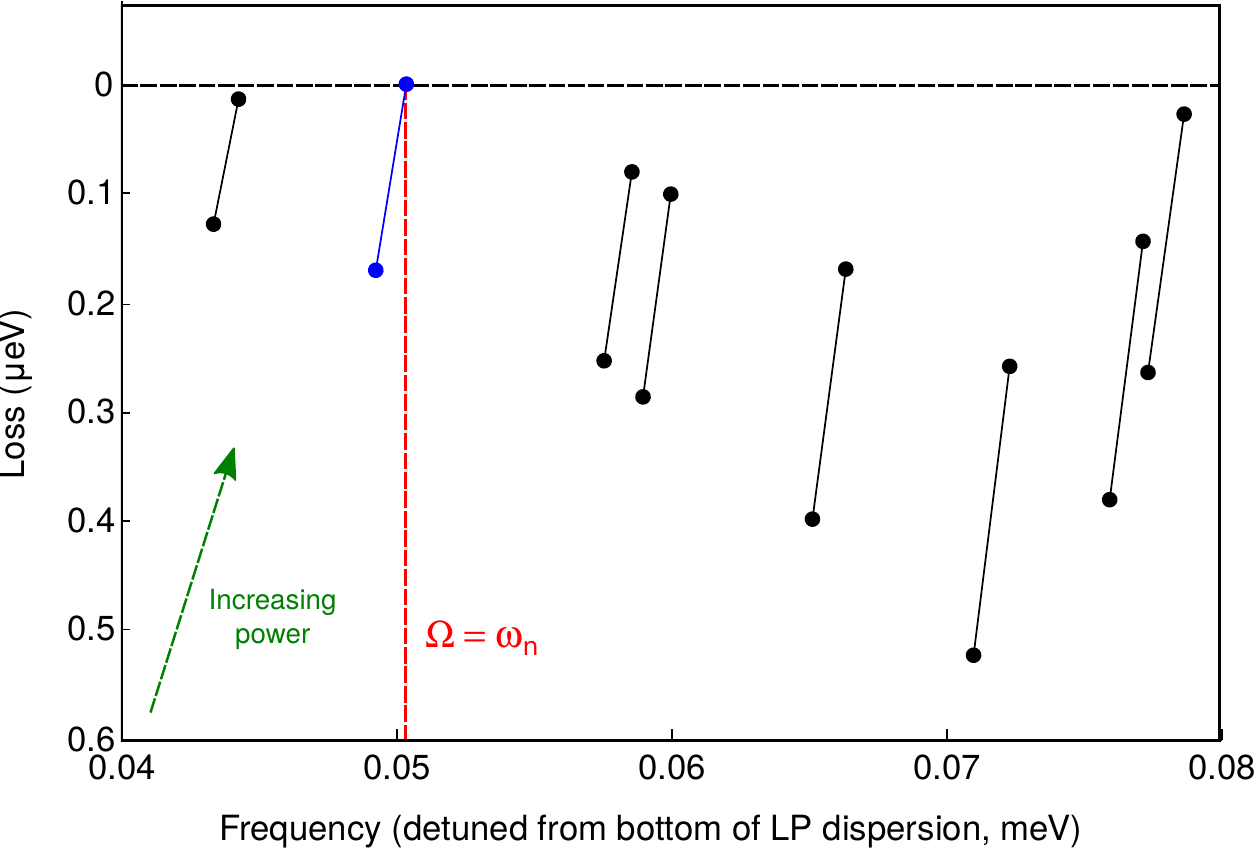}
\caption{\textbf{Linear threshold modes} Flow of (complex) eigenvalues of the linear non-Hermitian generator $\mathcal{H}_{\rm L}(P)$ as a function of pump power $P$ across the complex plane, computed for a fixed outgoing frequency $\Omega$. The flow direction is indicated by the arrow; eigenvalues approach the real axis from below as the pump power is increased. The lowest threshold mode is indicated in blue. It reaches the real line for the smallest pump power, \emph{and} has a real frequency $\omega_n$ equal to the imposed outgoing frequency $\Omega$.}
\label{complexFlow}
\end{center}
\end{figure}

\vspace*{0.3cm}
\noindent\textbf{Linear Threshold Modes.} We will now analyze steady-state condensate formation in the linearized regime. In particular, if we consider a single frequency steady-state ans\"atz for the condensate wavefunction:
\begin{align}
\Psi(\mathbf{r},t) = \varphi_n(\mathbf{r})e^{-i\omega_n t},
\end{align} 
the linearized GPE in Eq.~(\ref{linGPE}) of the main text becomes:
\begin{align}
\mathcal{H}_{\rm L}(P) \varphi_n(\mathbf{r}) = \left[ -\frac{\nabla^2}{2m} + \frac{g_R}{\gamma_R} Pf(\mathbf{r}) + \frac{i}{2} \frac{R}{\gamma_R} Pf(\mathbf{r}) - \frac{i}{2}\gamma_c \right] \varphi_n(\mathbf{r}) = \omega_n \varphi_n(\mathbf{r})
\end{align}
The condensate wavefunction for a single frequency $\omega_n$ condensate is therefore the $n$th eigenmode of the generator of linearized dynamics, $\mathcal{H}_{\rm L}(P)$. We require $\omega_n$ to be a purely real frequency for the steady-state solution to correspond to a nontrivial condensate mode; we will now discuss how this requirement determines the power threshold for a given spatial mode. For simplicity, we rewrite the above eigenproblem in the form:
\begin{align}
\left[ -\nabla^2 + sPf(\mathbf{r}) \right] \varphi_n(\mathbf{r}) = q^2\varphi_n(\mathbf{r})
\end{align}
where we have introduced the pump-induced potential $s$:
\begin{align}
\frac{s}{2m} = \frac{1}{\gamma_R} \left( g_R + \frac{i}{2} R \right)
\end{align}
and the `wavevector' $q(\omega_n)$ is defined by:
\begin{align}
\frac{q^2(\omega_n)}{2m} \equiv \omega_n + \frac{i}{2}\gamma_c
\end{align}
To determine the eigenmodes of $\mathcal{H}_{\rm L}(P)$, the above eigenproblem must be formulated as an appropriate boundary value problem (BVP); we make the following choice:
\begin{align}
\left[-\nabla^2 + sPf(\mathbf{r}) \right] \varphi_n(\mathbf{r}) &= q^2(\omega_n)\varphi_n(\mathbf{r})~,~\mathbf{r} \in \mathcal{P} \nonumber \\
-\nabla^2  \varphi_n(\mathbf{r}) &= q^2(\omega_n)\varphi_n(\mathbf{r})~,~\mathbf{r} \notin \mathcal{P}
\end{align}
where $\mathcal{P}$ is the region enclosed by the \emph{outer} edge $\partial\mathcal{P}$ of the pump, as defined in the main paper. Note here that we impose an `outgoing' boundary condition with wavevector $q(\omega_n)$ at the pump edge $\partial\mathcal{P}$, as opposed to the more usual case of considering a boundary far from the pump where the condensate wavefunction is vanishingly small and standard Dirichlet boundary conditions can be employed. For the large condensate sizes considered here, the latter approach would require simulating a very large spatial grid, making computation times inconveniently long. Our approach allows the use of a minimally relevant grid size. This occurs at a relatively minor expense: the outgoing wavevector imposed via this boundary condition depends on the unknown eigenvalue $\omega_n$, and this BVP therefore needs to be solved self-consistently. To do so, we fix the outgoing wavevector by choosing an outgoing frequency $\Omega$:
\begin{align}
\left[-\nabla^2 + sPf(\mathbf{r}) \right] \varphi_n(\mathbf{r}) &= q^2(\omega_n)\varphi_n(\mathbf{r})~,~\mathbf{r} \in \mathcal{P} \nonumber \\
-\nabla^2  \varphi_n(\mathbf{r}) &= q^2(\Omega)\varphi_n(\mathbf{r})~,~\mathbf{r} \notin \mathcal{P}
\end{align}
It is now a straightforward matter to solve this BVP for a range of (increasing) values of the pump power at a fixed $\Omega$; as a result, one obtains a set of eigenmodes $\{\varphi_n(\mathbf{r})\}$ and eigenfrequencies $\{\omega_n(P)\}$ of $\mathcal{H}_{\rm L}(P)$. These generally complex frequencies $\{\omega_n(P)\}$ flow across the complex plane as the pump-power is varied; an example of this flow is shown in Fig.~\ref{complexFlow}. For a certain pump power $P_n$, the $n$th eigenfrequency $\omega_n$ crosses the real axis (becomes real). The imaginary part of $\omega_n$ represents net loss, so its becoming zero implies that gain overcomes polariton loss at this pump power, and the associated eigenmode is an unstable fluctuation around the uncondensed polariton state. Furthermore, if the (now real) frequency is \emph{also} equal to the imposed outgoing frequency, that is $\omega_n = \Omega$, the wavevector $q(\omega_n)$ is equal both inside and outside the pump region $\mathcal{P}$. The self-consistency condition is therefore simultaneously fulfilled, and the corresponding $n$th eigenmode $\varphi_n(\mathbf{r}; \omega_n, P_n)$ represents a true condensate mode with real frequency $\omega_n$ and \emph{linearized} power threshold $P_n$. By varying the outgoing frequency $\Omega$, and computing eigenvalues as a function of pump power, a set of such linear threshold modes $\{\varphi_n(\mathbf{r}; \omega_n, P_n)\}$ can be obtained.

\vspace*{0.3cm}
\noindent\textbf{Continuity Equation and Linear Threshold Formula.} From the linearized dynamical equation for the condensate wavefunction, it is possible to obtain an equation for the dynamics of the condensate \emph{density}, $|\Psi|^2$. In particular,
\begin{align}
\frac{\partial |\Psi|^2}{\partial t} = \Psi^*\frac{\partial\Psi}{\partial t} + c.c.
\end{align}
From the generalized GPE (Eq.~(\ref{GPE})), it is easily found that:
\begin{align}
\Psi^*\frac{\partial \Psi}{\partial t} = \frac{i}{2m} \Psi^*\nabla^2\Psi + \left\{- ig_Rn_R -ig|\Psi|^2 + \frac{1}{2}\left(Rn_R - \gamma_c\right) \right\}|\Psi|^2
\end{align}
and so:
\begin{align}
\frac{\partial |\Psi|^2}{\partial t} = \frac{i}{2m} \left( \Psi^*\nabla^2\Psi - \Psi\nabla^2\Psi^* \right) + Rn_R|\Psi|^2 - \gamma_c|\Psi|^2
\end{align}
The first term on the right hand side has the form of the divergence of a probability current; this can be made more explicit by defining the probability current $\vec{j}$ as:
\begin{align}
\vec{j} = \frac{i}{2m} \left( \Psi \vec{\nabla}\Psi^* - c.c. \right)
\end{align}
following which the condensate density dynamics is governed by the equation:
\begin{align}
\frac{\partial |\Psi|^2}{\partial t} = Rn_R|\Psi|^2 -\nabla \cdot \vec{j} - \gamma_c|\Psi|^2
\end{align}
which has the well-defined form of a continuity equation. In particular, the above equation can be put into a more practical form by integrating over the area $\mathcal{P}$ of the region enclosed by the outer pump edge,
\begin{align}
\frac{\partial}{\partial t} \int_{\mathcal{P}} d^2\mathbf{r}~|\Psi|^2 = R\int_{\mathcal{P}}d^2\mathbf{r}~n_R|\Psi|^2 - \oint_{\partial \mathcal{P}} \vec{j} \cdot d\vec{s} - \gamma_c \int_{\mathcal{P}}d^2\mathbf{r}~|\Psi|^2
\end{align}
where the divergence theorem allows the term involving $\vec{j}$ to be rewritten as a flux integral. This equation has the simple interpretation: any increase in the total number of polaritons ($\propto \int_{\mathcal{P}}d^2\mathbf{r}~|\Psi|^2$) within the pump region comes from amplification via the exciton reservoir, at rate $R$. Losses to the polariton number can be attributed to either the mirror loss $\gamma_c$, or a leakage of the condensate from the pump edge. Since we are integrating within the \emph{outer} pump edge $\partial \mathcal{P}$, beyond which by definition no source of polariton production exists, there can be no incoming probability current that would increase the polariton number within the pump region.

Now, we narrow our focus to the linearized regime, where the reservoir density $n_R = Pf(\mathbf{r})/\gamma_R$ as shown earlier. Furthermore, we consider a single mode solution such that $\Psi(\mathbf{r},t) = \varphi_n(\mathbf{r}; \omega_n, P_n)e^{-i\omega_nt}$, where $\varphi_n(\mathbf{r}; \omega_n, P_n)$ is the eigenmode of $\mathcal{H}_{\rm L}(P_n)$ that has (real) eigenfrequency $\omega_n$. For simplicity, we suppress the parameters defining $\varphi_n$ in what follows. With this ans\"atz, the condensate density is time-independent and the above continuity equation reduces to:
\begin{align}
\frac{R}{\gamma_R}P \int_{\mathcal{P}}d^2\mathbf{r}~f(\mathbf{r})|\varphi_n|^2 = \oint_{\partial \mathcal{P}} \vec{j}[\varphi_n] \cdot d\vec{s} + \gamma_c \int_{\mathcal{P}}d^2\mathbf{r}~|\varphi_n|^2
\end{align}
Here, the probability current $\vec{j}[\varphi_n]$ is now evaluated for the eigenmode $\varphi_n$, as in the main paper. Now, defining the condensate density $\rho_n$, pump overlap $\mathcal{G}_n$, and in-plane loss $\gamma_n$ respectively as in the main paper:
\begin{align}
\rho_n = \int_{\mathcal{P}}d^2\mathbf{r}~|\varphi_n|^2~,~\mathcal{G}_n =  \frac{1}{\rho_n}\int_{\mathcal{P}}d^2\mathbf{r}~f(\mathbf{r})|\varphi_n|^2~,~\gamma_n = \oint_{\partial \mathcal{P}} \vec{j}[\varphi_n] \cdot d\vec{s},
\end{align}
we can recover the linear threshold formula (Eq.~(\ref{thresholdFormula}) of the main text):
\begin{align}
\frac{P_n}{P_0} = \frac{1+ \gamma_n/(\rho_n\gamma_c)}{\mathcal{G}_n} \equiv \frac{1+\Gamma_n}{\mathcal{G}_n}
\end{align}
with $P_n$ being the linear threshold power for the $n$th mode, and $P_0 = (\gamma_c\gamma_R)/R$.

\begin{figure}[dthp]
\begin{center}
\includegraphics[scale=0.6]{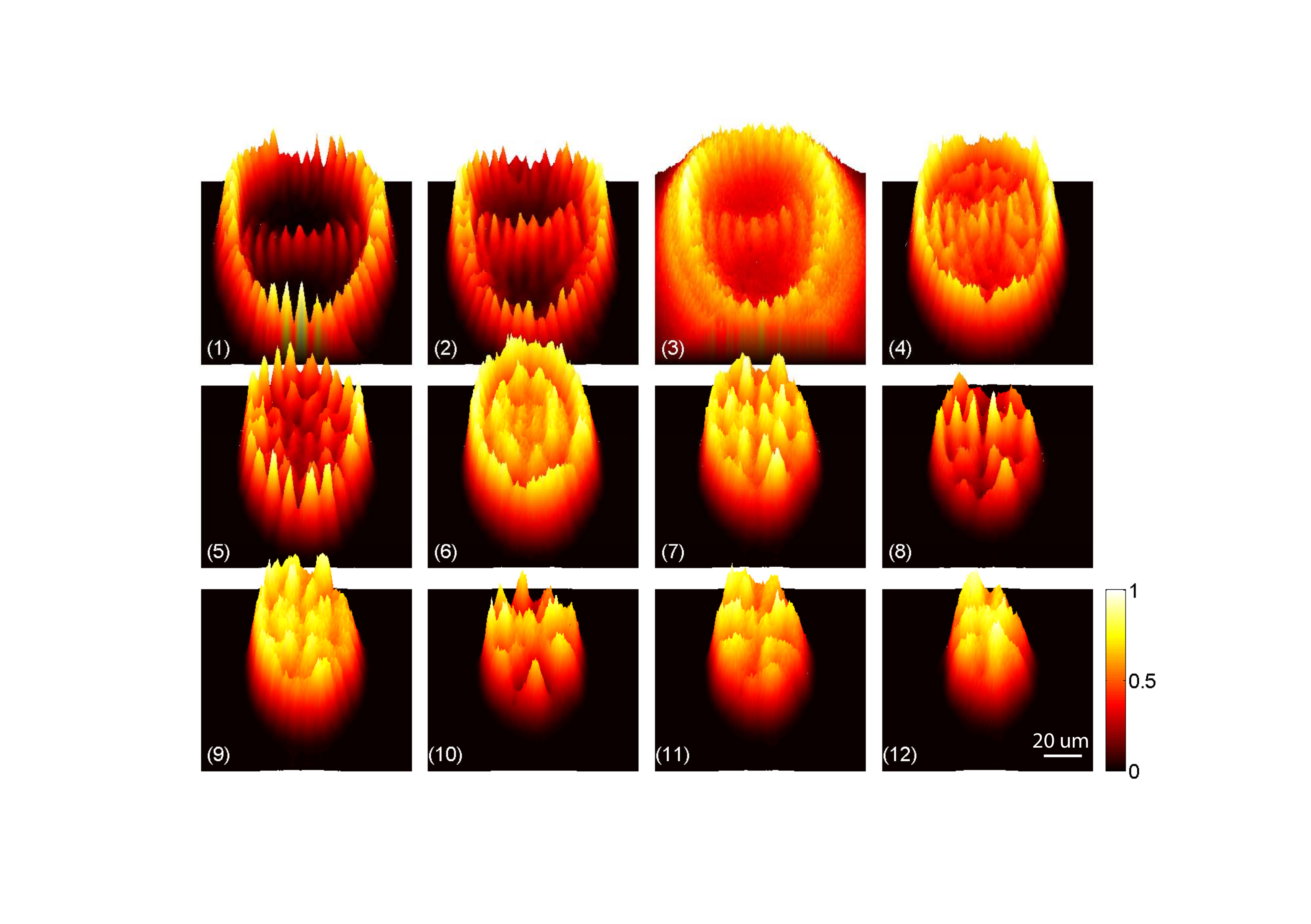}
\caption{(1)-(12) Mixed modes for selected points in the phase diagram plot shown in Fig.~S6. The scale bar in (12) indicates  20 $\mu$m.}
\label{mixedmodes}
\end{center}
\end{figure}

\begin{figure}[t]
\begin{center}
\includegraphics[scale=0.6]{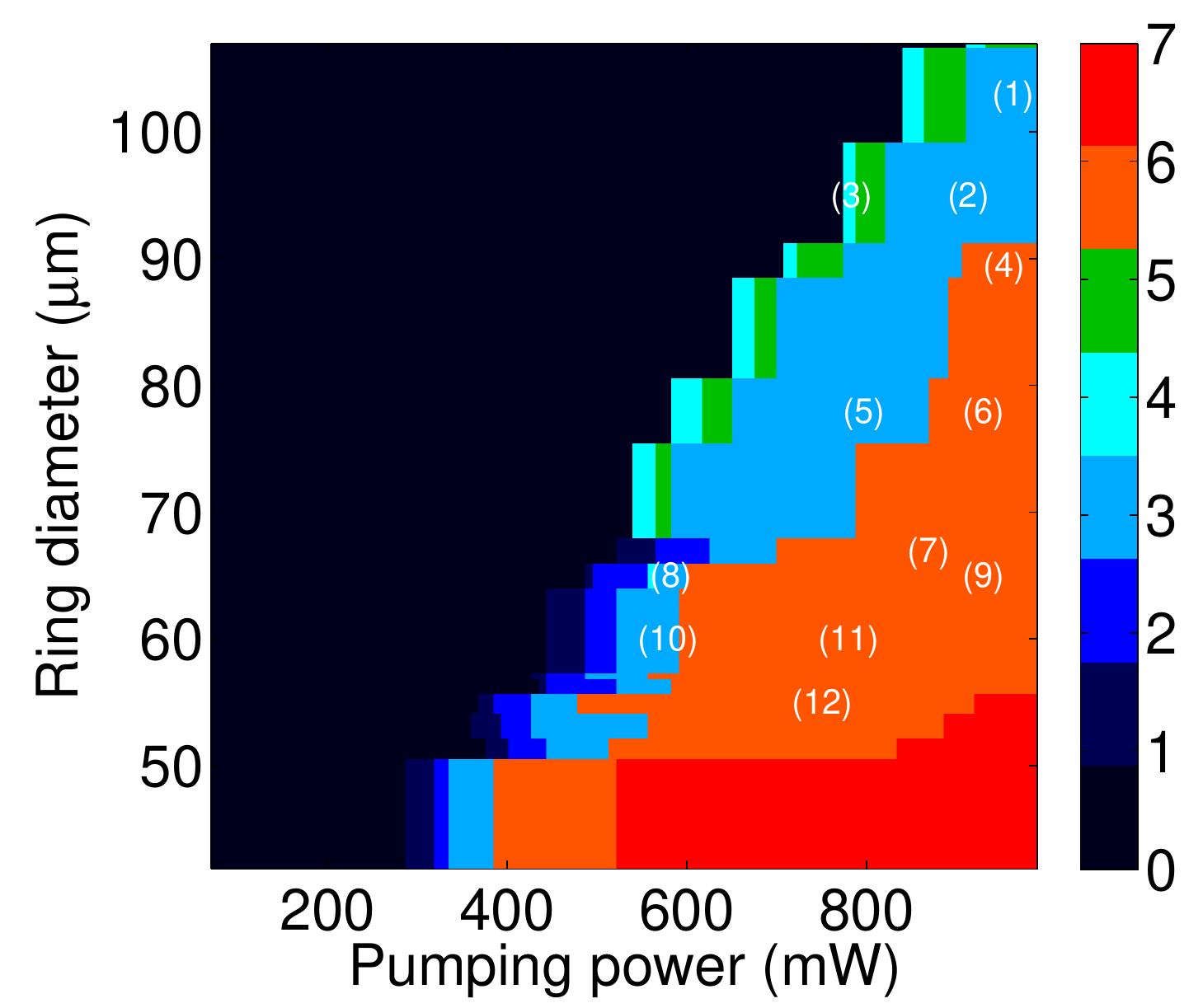}
\caption{Phase diagram indicating the position of the modes whose spatial distributions are shown in Fig.~S5.}
\label{phasepos}
\end{center}
\end{figure}

\vspace*{0.3cm}
\noindent\textbf{Spatial distributions of mixed modes in the optical trap.} Because of the interactions among high-order modes, a large set of mixed modes shows up in the phase diagram. In Fig.~\ref{mixedmodes}, we plot the spatial distributions for 12 mixed modes whose positions in the phase diagram are marked in Fig.~\ref{phasepos}. As can be seen, modes (1)-(3) have 
both ripple and petal characteristics, and modes (4)-(7) and (9)-(11) are petal-like and are quantized in the azimuthal direction, although the emission intensities from peaks and nodes are comparable.  Modes (8) and (12) are ripple-like. Based on our numerical simulations, mixed modes are a direct consequence of interactions between high-order modes with very close thresholds, rather than being artifacts from time-integrated measurements.

\begin{figure}[thpd]
\begin{center}
\includegraphics[scale=0.6]{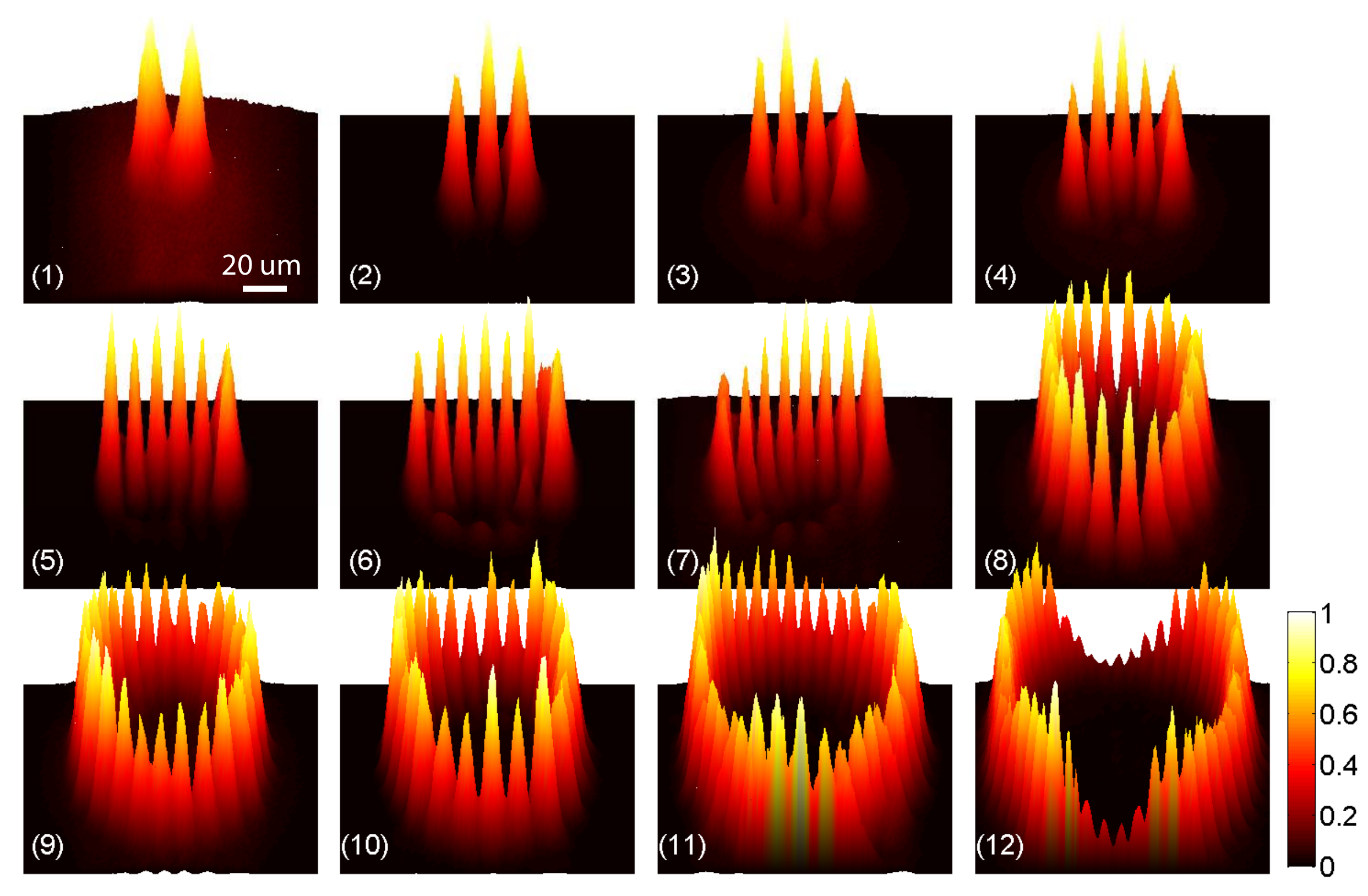}
\caption{(1)-(12) Spatial distributions of ripple and petal states with different number of nodes when the pumm parameters are varied.  The scale bar in (1) indicates  20 $\mu$m.}
\label{puremodes}
\end{center}
\end{figure}

\vspace*{0.3cm}
\noindent\textbf{Evolution of node numbers of ripple and petal states  in the optical trap.} The number of nodes in either ripple or petal states can be varied by adjusting pump diameter. The pump power also needs to be increased in order to reach condensation threshold as the diameter of the pump increases. In Fig.~\ref{puremodes}, we show the spatial distributions of 12 distinct modes with various nodes.  As can be seen, by changing the pump diameter, we could switch from a 2-node ripple state up to 8-node ripple states continuously. Similar switching behavior can also be realized by using petals as shown in (8)-(12).

\end{document}